\documentclass[fleqn,usenatbib]{mnras}

\usepackage{newtxtext,newtxmath}

\usepackage[T1]{fontenc}
\usepackage{ae,aecompl}

\usepackage{graphicx}	
\usepackage{amsmath}	
\usepackage{amssymb}	

\newcommand\cm{{\rm\thinspace cm}}

\newcommand\erg{{\rm\thinspace erg}}

\newcommand\g{{\rm\thinspace g}}
\newcommand\keV{{\rm\thinspace keV}}
\newcommand\km{{\rm\thinspace km}}
\newcommand\kpc{{\rm\thinspace kpc}}

\newcommand\Msun{\hbox{$\rm\thinspace M_{\odot}$}}

\newcommand\s{{\rm\thinspace s}}

\newcommand\cmsq{\hbox{$\cm^2\,$}}

\newcommand\ergpcmsqps{\hbox{$\erg\cm^{-2}\s^{-1}\,$}}

\newcommand\ergps{\hbox{$\erg\s^{-1}\,$}}

\newcommand\gps{\hbox{$\g\s^{-1}\,$}}

\newcommand\kmps{\hbox{$\km\s^{-1}\,$}}

\newcommand\psqcm{\hbox{$\cm^{-2}\,$}}

\newcommand\rg{\hbox{$r_{\rm g}\,$}}

\title[Emission from the Plunge Region?]{The Soft State of the Black
  Hole Transient Source MAXI\,J1820+070: Emission from the Edge of the Plunge Region?} \author[A.C. Fabian et al.] {\parbox[]{6.5in}{{A,C.~Fabian$^1\thanks{E-mail: acf@ast.cam.ac.uk}$,
      D.J~Buisson$^1$, P.~Kosec$^1$, C.S.~Reynolds$^1$,   
D.R.~Wilkins$^2$, J.A.~Tomsick$^3$, D.J.~Walton$^1$,  P. Gandhi$^4$, D. Altamirano$^4$, Z.~Arzoumanian$^5$ E.M.~Cackett$^6$, S. Dyda$^1$, J.A. Garcia $^7$, K.C.~ Gendreau$^5$, B.W~Grefenstette$^7$, F.A.~Harrison$^7$, J. Homan$^{8,9}$, E. Kara$^{10}$, R.M.~ Ludlam$^{7}$, J.M.~Miller$^{11}$ and J.F. Steiner$^{12}$}\\
    \footnotesize
    $^1$ Institute of Astronomy, Madingley Road, Cambridge CB3 0HA\\
$^2$ Kavli Institute for Particle Astrophysics and Cosmology, Stanford University, 452 Lomita Mall, Stanford, CA 94305, USA\\
$^3$ Space Sciences Laboratory, 7 Gauss Way, University of California, Berkeley, CA 94720-7450, USA\\
$^4$ Department of Physics \& Astronomy, University of Southampton, Highfield, Southampton SO17 1BJ, UK\\
$^5$ Astrophysics Science Division, NASA?s Goddard Space Flight Center, Greenbelt, MD 20771, USA \\
$^6$ Department of Physics \& Astronomy, Wayne State University, 666 W. Hancock Street, Detroit, MI 48201, USA\\
$^7$ Cahill Center for Astronomy and Astrophysics, California Institute of Technology, Pasadena, CA 91125, USA\\
$^8$ Eureka Scientific, Inc., 2452 Delmer Street, Oakland, CA 94602, USA\\
$^9$ SRON, Netherlands Institute for Space Research, Sorbonnelaan 2, 3584 CA Utrecht, The Netherlands\\
$^{10}$ MIT Kavli Institute for Astrophysics and Space Research, 70 Vassar Street, Cambridge, MA 02139, USA\\
$^{11}$ Department of Astronomy, University of Michigan, 1085 South University Avenue, Ann Arbor, MI 48109-1107, USA\\
$^{12}$ Center for Astrophysics, Harvard University, 60 Garden Street, Cambridge, MA 02138, USA\\
}}
\date{Accepted XXX. Received YYY; in original form ZZZ}

\pubyear{2019}

\begin{document}
\label{firstpage}
\pagerange{\pageref{firstpage}--\pageref{lastpage}}
\maketitle

\begin{abstract}
The Galactic black hole X-ray binary MAXI J1820+070 had a bright outburst in 2018 when it became the second brightest X-ray source in the Sky. It was too bright for X-ray CCD instruments such as {\it XMM-Newton} and {\it Chandra}, but was well observed by photon-counting instruments such as {\it NICER} and {\em  NuSTAR}. We report here on the discovery of an excess emission component  during the soft state. It is best modelled with a blackbody spectrum in addition to the regular disk emission, modelled either as \texttt{diskbb} or \texttt{kerrbb}.  Its temperature varies from about 0.9 to 1.1 keV which is about 30 to 80 per cent higher than the inner disc temperature of \texttt{diskbb}.  Its flux varies between 4 and 12 percent of the disc flux. 
Simulations of magnetised accretion discs have predicted the possibility of  excess emission associated with a non-zero torque at the Innermost Stable Circular Orbit (ISCO) about the black hole, which from other {\em NuSTAR} studies lies at about 5 gravitational radii or about 60 km (for a  black hole mass is $8\Msun$). In this case the emitting region at the ISCO has a width varying between 1.3 and  4.6 km and would encompass the start of the plunge region where matter begins to fall freely into the black hole. 
 \end{abstract}


\section{Introduction}
On 2018 March 6, Tucker et al. (2018) discovered a new optical source from the 
All-Sky Automated Survey for SuperNovae project (Shappee
et al. al. 2014) and designated it ASASSN-18ey.  5 days later
the Monitor of All-sky X-ray Image (MAXI; Matsuoka et al. 2009) on the
International Space Station (ISS) detected its X-ray emission
(Kawamuro et al. 2018) and classified it as a new Galactic black hole
candidate MAXI J1820+070 (see Shidatsu et al. 2019 for a summary of
observations, in particular of those made by MAXI). The source
displayed a hard spectrum which rapidly rose to a peak flux making it
the second brightest object in the X-ray Sky. It then slowly decayed
in brightness while remaining in the hard state before, in 2018 June,
the spectrum became soft, peaking in flux before slowly decaying
again (Fig.~\ref{fig:bat_lc}). Later in September 2018 it turned hard and continued to drop in
flux.  The hard-soft-hard behaviour is typical for a stellar mass
black hole binary source. 
The black hole nature of the system has recently been confirmed by optical spectroscopy (Torres et al. 2019). 

X-ray detectors on the Neutron star Inner Composition Explorer ({\em NICER})
instrument (Gendreau et al. 2016), also on the ISS, recorded incident count rates up
to 50,000 counts $\s^{-1}$ in the hard state. When at  20,000 counts $\s^{-1}$, Kara et al.
(2019) found  X-ray reverberation lags of a msec and less,
confirming its status as an accreting black hole. The amplitude of the
lag changed systematically during the decay from the first peak
indicating that the corona was shrinking in height. This agrees with the behaviour of 
X-ray spectra from the Nuclear Spectroscopic Telescope Array
({\em NuSTAR}; Buisson et al. 2019).

Here we study {\em  NuSTAR} and {\em NICER} spectra from the Soft Sate of MAXI
J1820+070. We use a {\em NICER} spectrum to confirm the disk blackbody
nature of the emission in the energy range 0.8-7 keV and {\em  NuSTAR} spectra from several
epochs in the soft state to examine the spectrum from 3--78
keV. Significant excess emission found in the energy range 6--10 keV
is best modelled by an additional blackbody component. It is plausibly
explained as emission from the transition between the stably orbiting
accretion disc and the plunge region, where matter spirals into the
black hole (Hawley \& Krolik 2002; Machida \& Matsumoto 2003, Zhu et al.
2012).

\section{The X-ray Spectra}

\begin{figure}
  \includegraphics[width=\columnwidth]{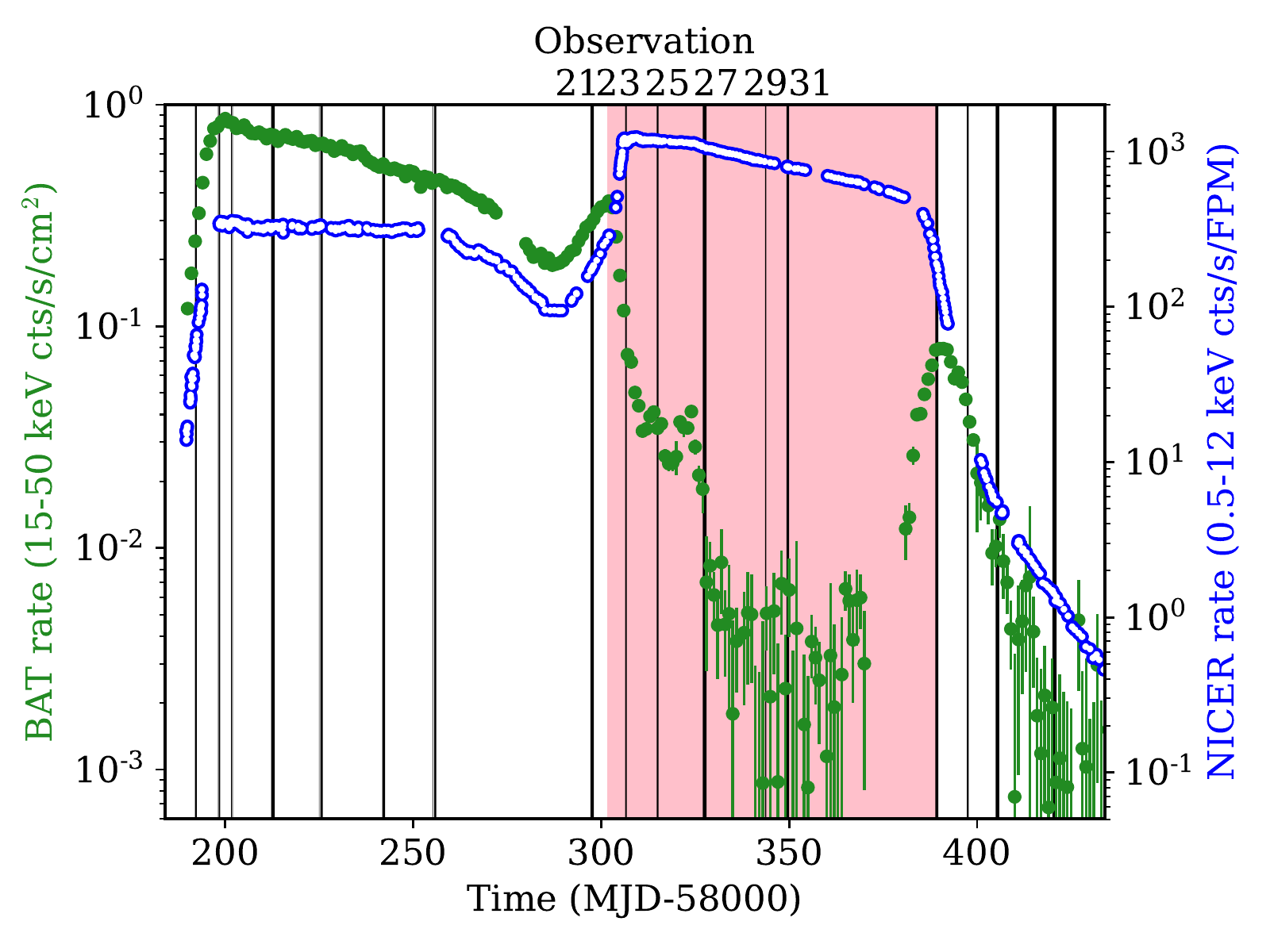}
  \caption{Swift-BAT and NICER lightcurves with {\em NuSTAR} observations discussed here  indicated by vertical lines. The pink region covers the soft state. The start dates and observation length and live fractions of observations 21 through 31 are 28 June (77.7 ks, 0.5), 7 July (38.1, 0.33), 15 July (43.7, 0.39), 28 July (83.3, 0.45), 13 Aug (26.5, 0.54) 19 Aug (58.7,0.58).
  }
  \label{fig:bat_lc}
\end{figure}

We concentrate on 5 soft state {\em  NuSTAR} spectra, here designated as Nu23, 25, 27, 29 and 31 (they are distinguished by the last two digits in their OBSIDs). Monitoring
of the source by the hard X-ray Burst Alert Telescope (BAT) on the Neil Gehrels
Swift satellite show minimum flux from 2019 July 28 and Sept 10, during which 3 {\em  NuSTAR} spectra were obtained, Nu27 through 31
respectively (Fig.~\ref{fig:bat_lc}).  A 1.5\,ks {\em NICER} spectrum from July is contemporaneous with  
Nu29.  {\em  NuSTAR} spectra have been reduced as described in Buisson et al. 2019). 
The {\em NICER} spectrum has been analysed with response matrix version 1.02 corrected by application of a function
derived from observations of the Crab nebula (Ludlam et al. 2018). A radio flare occurred just before observation Nu23 (Bright et al. 2018), which can be considered as in a short intermediate state. 

\begin{figure}
  \includegraphics[width=1.\columnwidth]{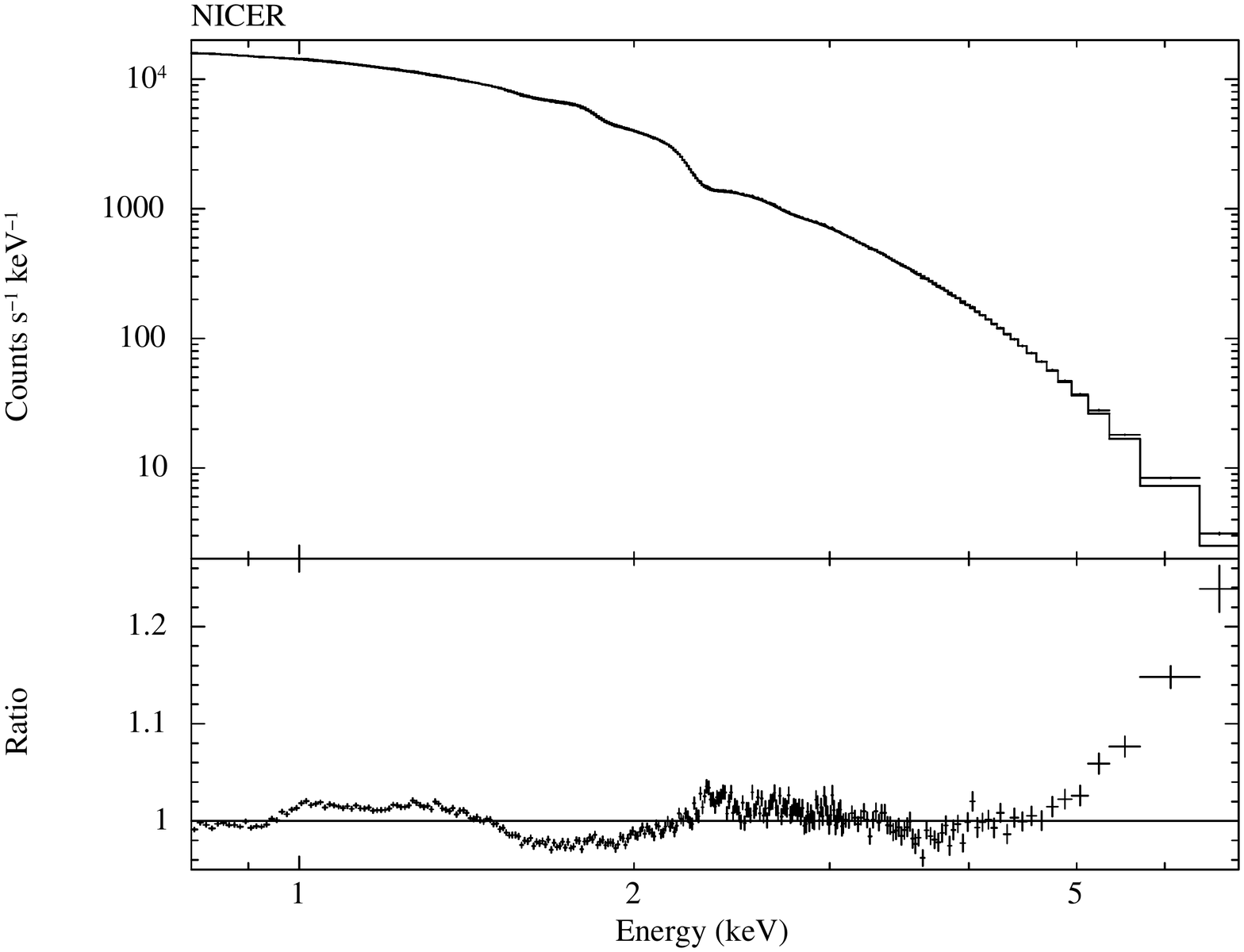}
  \includegraphics[width=1.\columnwidth]{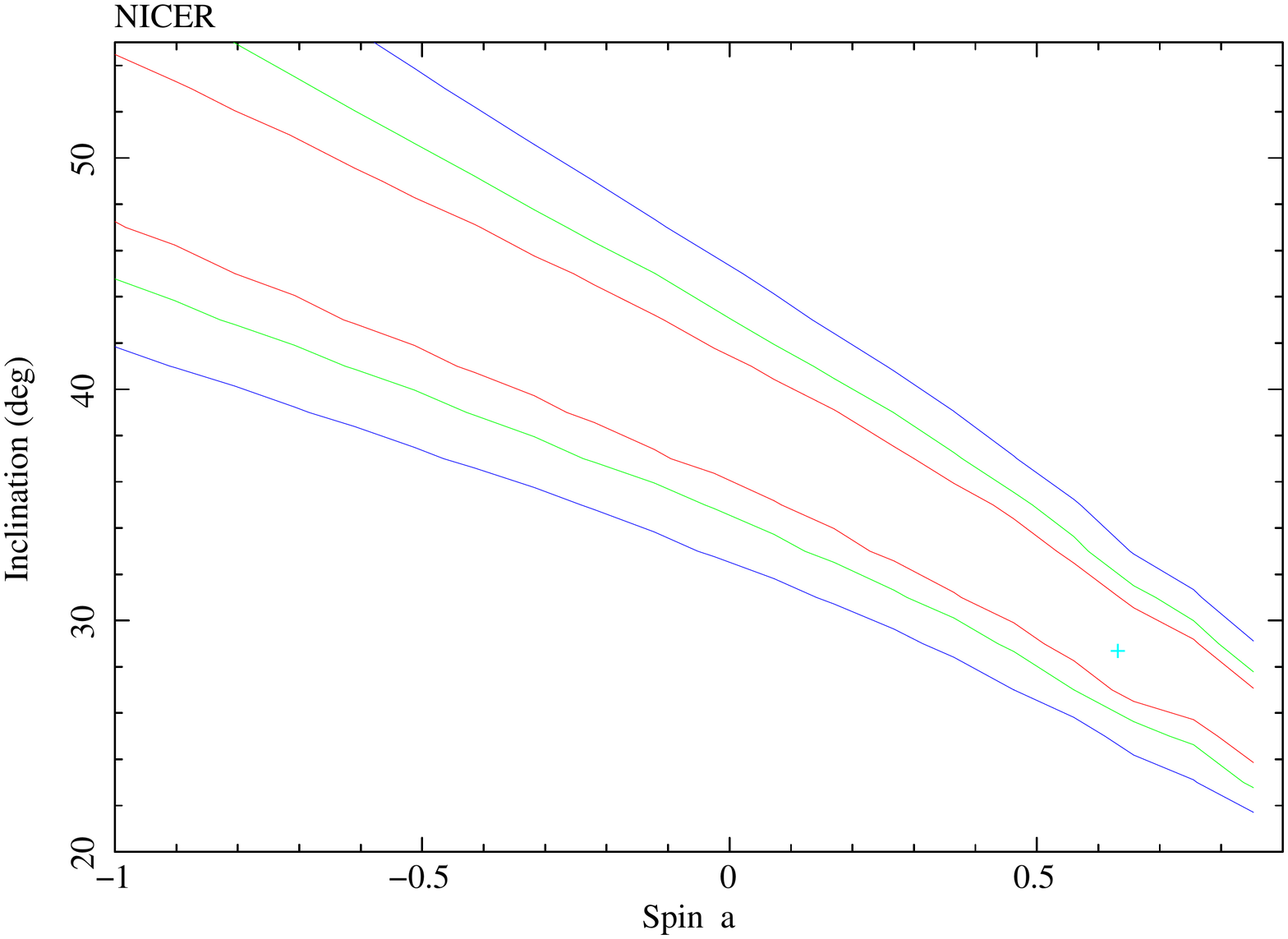}  
  \caption{Top: {\em NICER} spectrum from {\em  NuSTAR} epoch 29. Most of the residuals from 0.8 to 3 keV are instrumental. Bottom: Allowed region using \texttt{kerrbb}. The hard state results from Buisson et al. (2019) indicate that the spin lies between about 0 and 0.5 and inclination between 30 and 40 deg, which is compatible with this result.}
  \label{fig:nikbbfit_nicon}
\end{figure}

\begin{figure}
  \includegraphics[width=1\columnwidth]{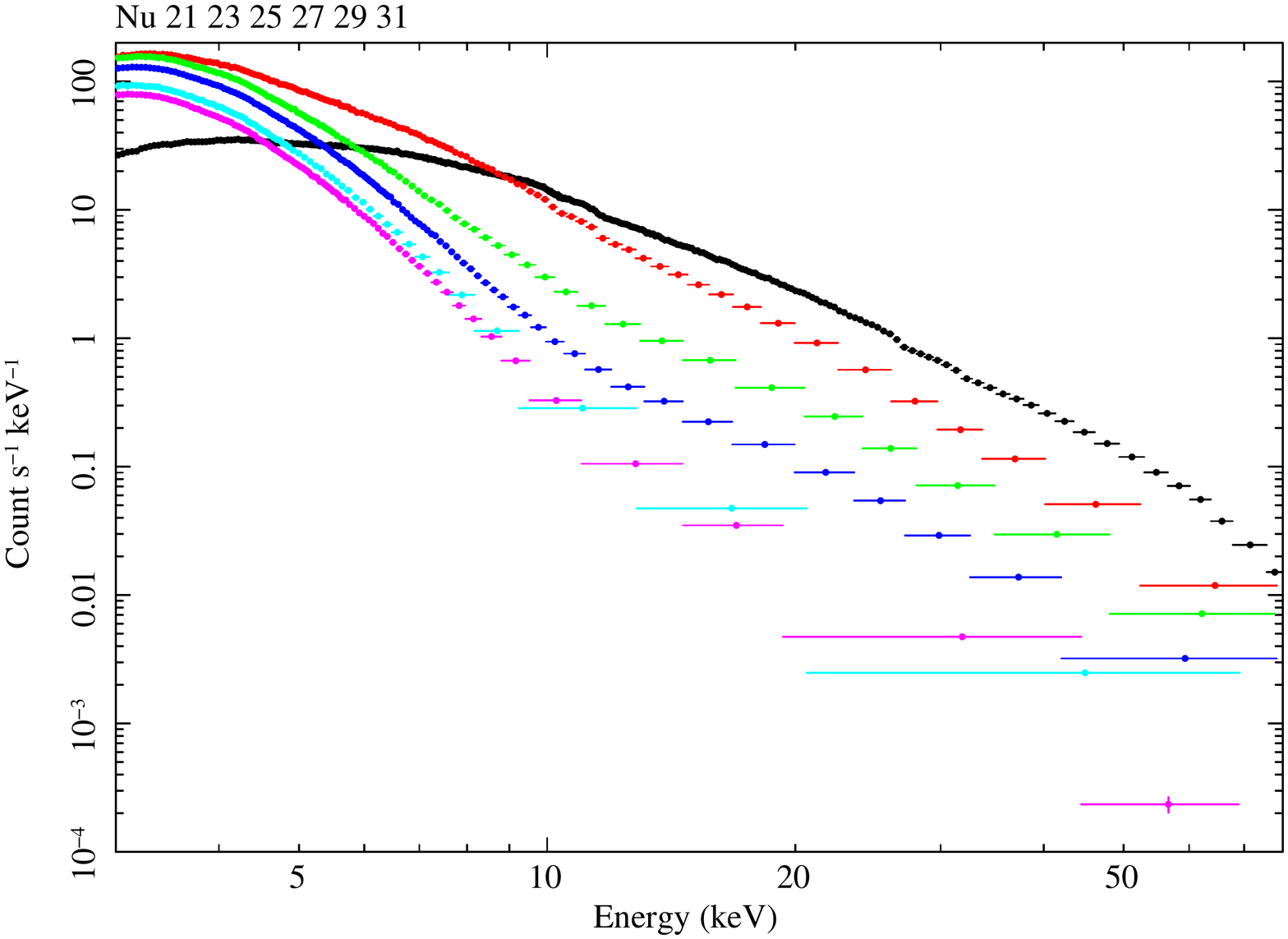} 
  \includegraphics[width=1\columnwidth]{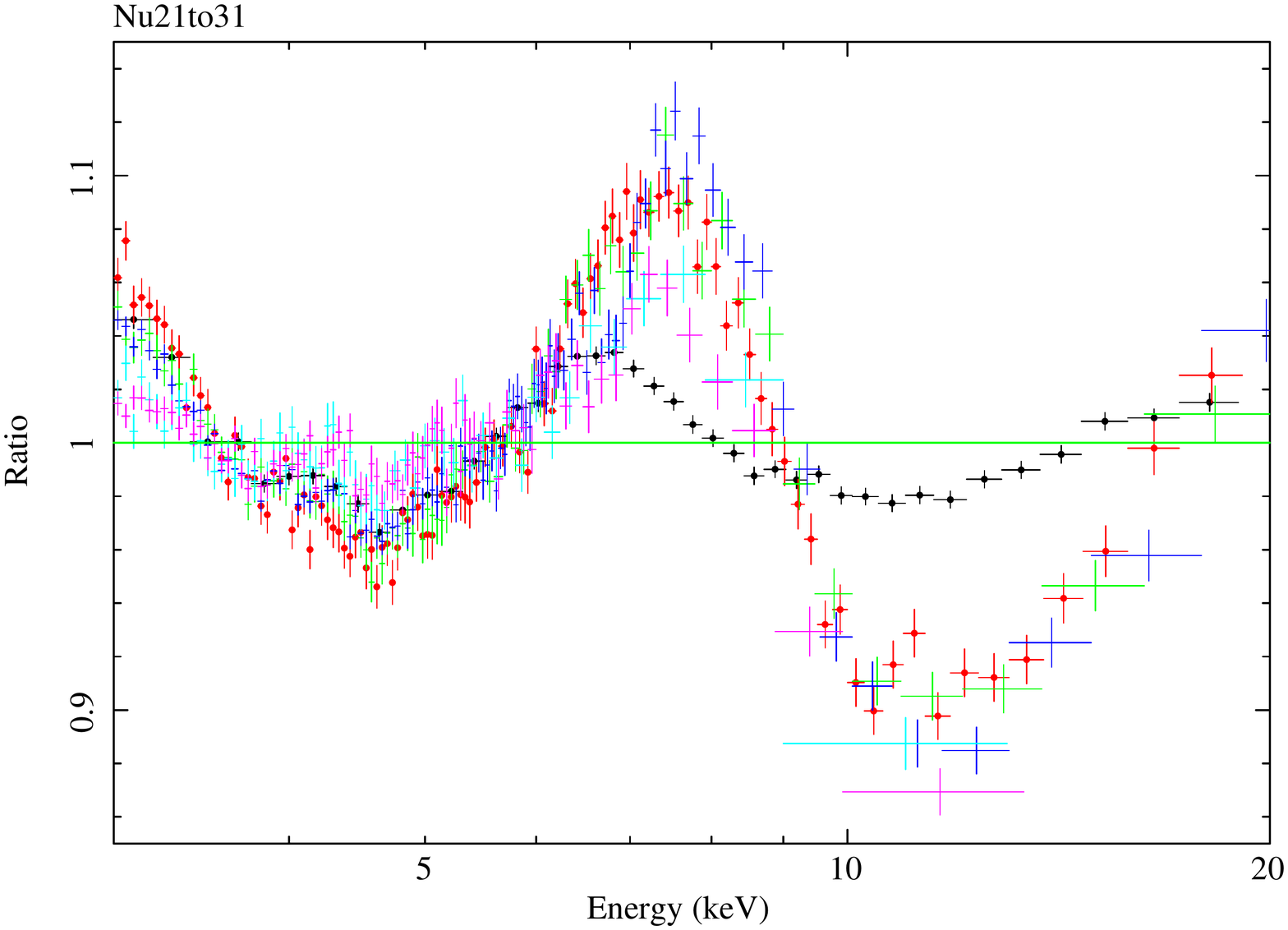}  
 \caption{Top: {\em  NuSTAR} module A spectra of observations 23 through 31 with ratios to the best-fitting powerlaw below. The black points are from observation 21 which is on the flux rise to the soft state.  Note that the black, hard-state residuals peak at 6.5 keV whereas those in the soft state peak around 7.5 keV.   Note that the underlying continuum, which forms the denominator of the ratios plotted here, falls very rapidly over the energy band shown. }
 \label{fig:nu2131spec_rat_ratbin}
\end{figure}

\begin{figure}
  \includegraphics[width=1\columnwidth]{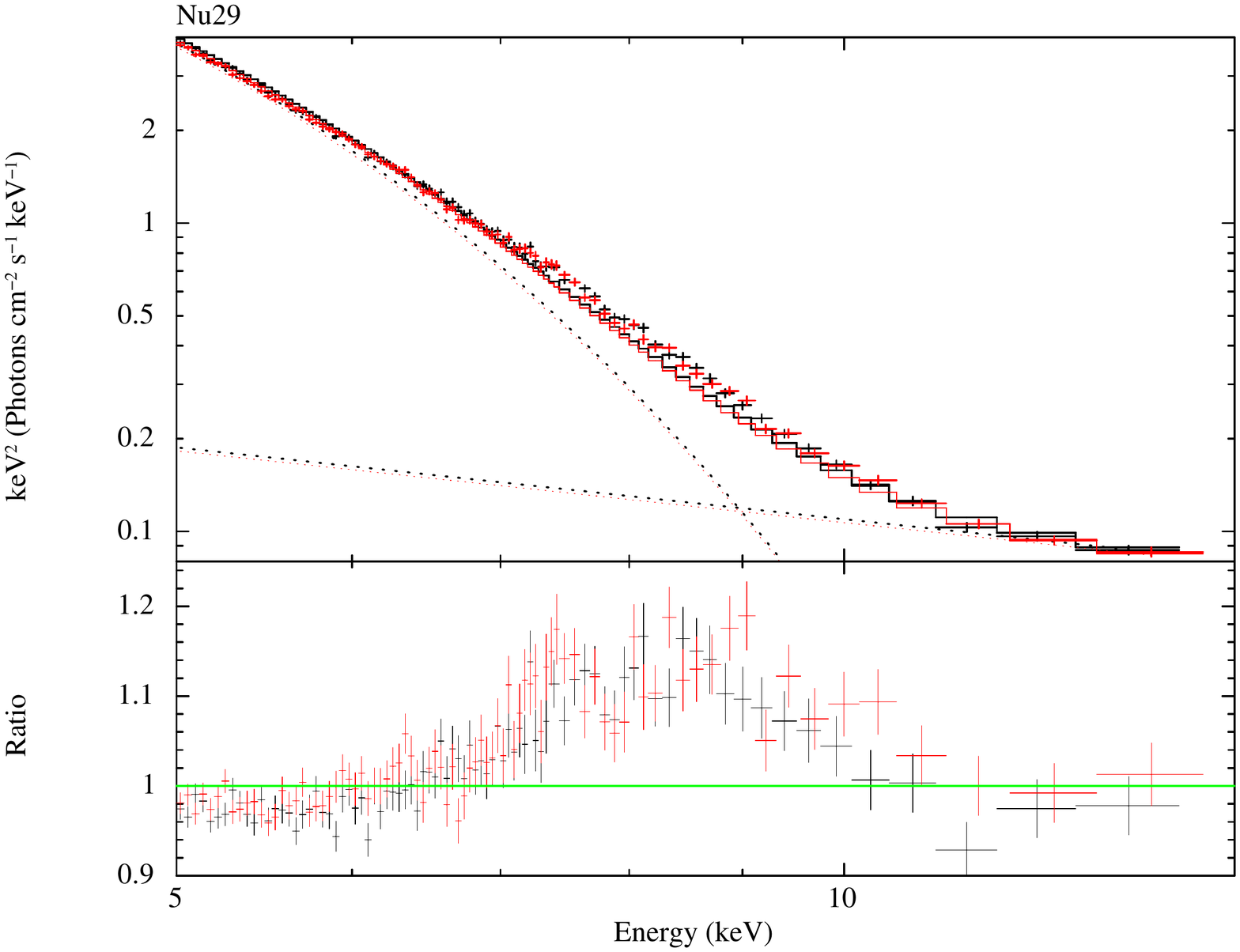}   
   \caption{ {\em  NuSTAR} unfolded spectrum from observation 29 showing the subtle nature of the excess emission in the Wien tail of kerrbb. Here a power-law continuum has been fitted to the data above 10 keV and is only for illustrative purposes as it is not the best fit in a $\chi^2$ sense, which prefers a dip at 10 keV as seen in Fig. 4. } 
   \label{fig:nu29rat}
\end{figure}

\subsection{A {\em NICER} soft-state spectrum}

We fit the {\em NICER} spectrum (OBSID 1200120236) with a simple absorption model \texttt{TBNEW} applied to a
\texttt{kerrbb} model, which is a fully relativistic representation of
the quasi-blackbody emission expected from a thermal accretion disc (Li
et al. 2005). The distance to the source is set at 3.5 kpc, as
determined from Gaia parallax data (Gandhi et al. 2019) who derive a distance of
$3.5 \pm 1.5\kpc$).  The mass function of the binary system has been determined from optical spectroscopy by Torres et al. (2019) to be $f(M)=5.18\pm0.15\Msun$. They obtained an orbital inclination of between 69 and 77 deg from the lack of a continuum eclipse and periodic changes in H$\alpha$ equivalent width.  This implies a black hole mass of $7-8\Msun$.  Buisson et al. (2019) found  the
inclination of the inner accretion disc to lie between 30 and 40 deg from relativistic reflection
spectral fits to the {\em  NuSTAR} hard state spectra. 

Our  best fit is
shown in Fig.~\ref{fig:nikbbfit_nicon} with the constraints on spin and inclination in the lower panel. Reduced $\chi^2$ is 8 over the 0.8 to 7 keV range arises from known calibration inadequacies in the response files used here. Adding systematics of
1.5\ per cent brings the reduced $\chi^2$ close to unity (0.99). The level of absorption indicates a low column density of $N_{\rm H}\sim 5 \times 10^{20}\psqcm$ and
the black hole mass implied by the \texttt{kerrbb} model lies between
$5-10\Msun$. We note an excess in the spectrum above 6 keV. If the inclination of the inner disc is $\sim70$\,deg, to match the optically-derived orbital inclination, then we find that the spin must be close to maximally retrograde, i.e. $a<-0.95$. If however the innerdisc inclination is lower at 30--40\,deg, then the spin lies approximately between $\pm0.5$.    

 More detailed spectral fits involving the hard  state and both {\em NICER} and {\em  NuSTAR} will be reported elsewhere. 

\subsection{The {\em  NuSTAR} Spectra}

\subsubsection{Phenomenology and the Need for an Extra Blackbody Component}

We then examine the five sets of {\em  NuSTAR} observations that span the soft state, here designated Nu23, Nu25, Nu27, N29 and Nu31.  For comparison, we add Nu21 which is in the second rise of the hard state before the abrupt drop in  BAT 14-195 keV count rate marks the beginning of the soft state. Nu23 and Nu25 have significant BAT flux, whereas Nu27 through Nu31 are faint in the BAT. The count spectra of {\em  NuSTAR} module A\footnote{Just one module is used here for illustrative purposes to avoid cluttering the figure too much. Both Module A and B are used in all spectral parameter fits.} are plotted in \ref{fig:nu2131spec_rat_ratbin} with the ratios of those spectra to a freely-fitted \texttt{diskbb + po} model  are shown below. No absorption is used as its effect is negligible at the energies of the {\em NuSTAR} spectra. The ratio spectrum of Nu21 (shown in black markers in Fig.~\ref{fig:nu2131spec_rat_ratbin}) is typical of the hard state with a  peak centred at about 6.5 keV which corresponds to the broad iron line (Buisson et al. 2019). Spectral fits to the spectrum of Nu21 with a double corona model (epoch 8 in Buisson et al. 2019) reveal that the weak upper corona in Nu21 lies at a height of at least $100r_{\rm g}$ and the coronal temperature is also high at $kT\sim 230 \keV$, the bulk of the power-law is emitted close to the black hole ($<10r_{\rm g}$). The disc blackbody component is detectable only below 4 keV.

In contrast, the soft-state ratio spectra Nu23 -- Nu31 (red through magenta in Fig.~\ref{fig:nu2131spec_rat_ratbin}) are dominated by the disc blackbody spectrum up to about 10 keV. This is due principally to  the disc emission becoming much stronger but also to the powerlaw component becoming progressively weaker. There is again a peak in the ratio spectra but now it is much stronger and its peak shifts to almost 8  keV by Nu27 (Fig.~\ref{fig:nu2131spec_rat_ratbin})

The excess occurs where the Wien tail of the disc blackbody emission is steepest as shown in Fig.~\ref{fig:nu29rat}.
A further  power-law to account for this has been mimicked by the {\sc SIMPL} model (Steiner et al. 2009), but it needs to be implausibly steep with a photon index  $\Gamma=7-8$.  We have also briefly considered radiative transfer effects such as those responsible for the spectral correction factor commonly applied to accretion disc models. The emission is at too high an energy to be related to changes in free-free absorption within the disc.

\begin{figure} 
  \includegraphics[width=1.\columnwidth]{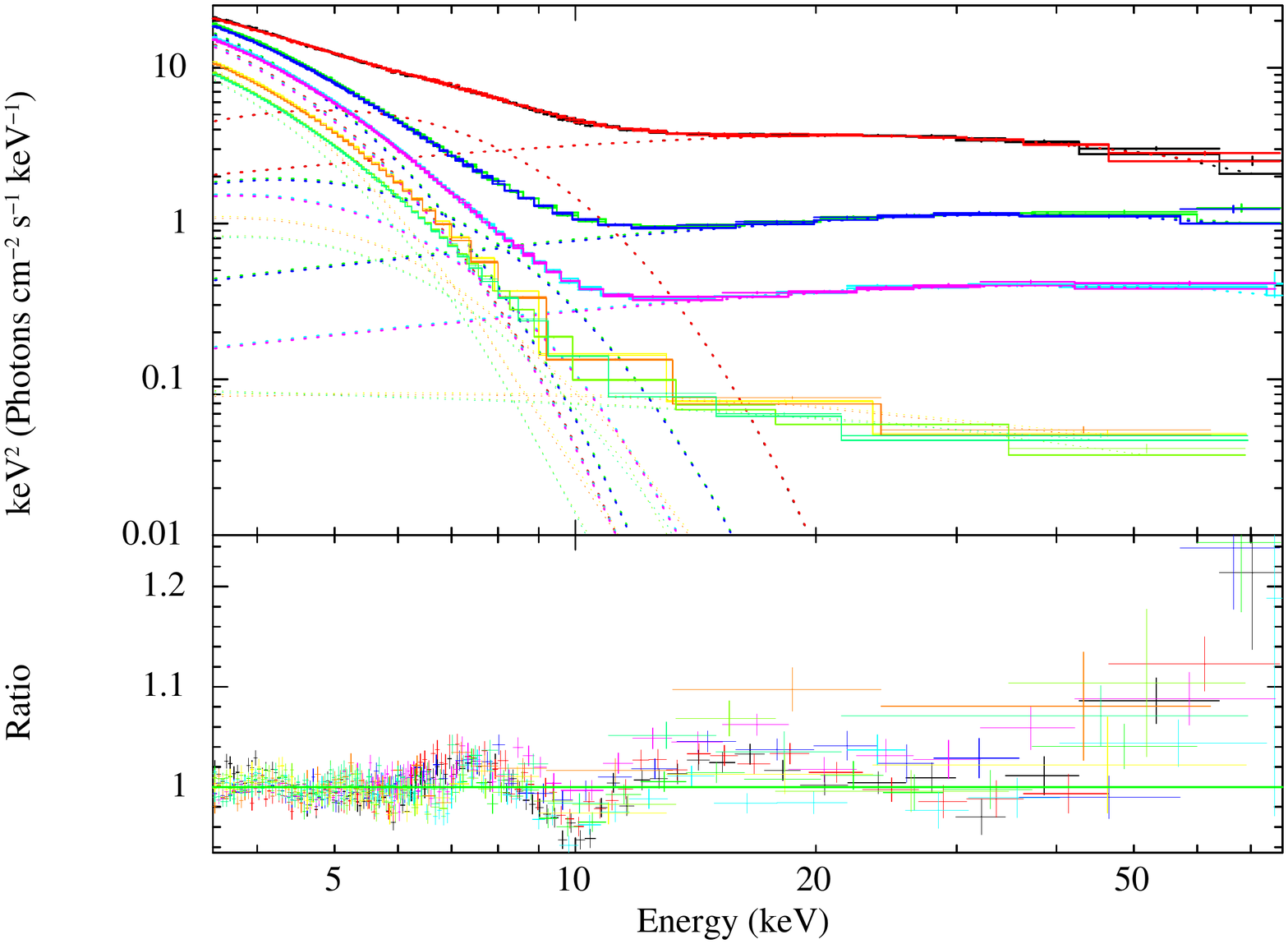}
\caption{Top: Unfolded spectra of Nu23 to 31 using  \texttt{kerrbb} with an extra blackbody and cutoff power-law components. The parameters are given in Table 1.}
\label{fig:nu2331dbb_kbb}
\end{figure}

 \begin{table*}\centering
  \caption{The best-fit model parameters for the model
    \texttt{diskbb+bb+cutoffpl}.
    Errors are quoted at the 90
per cent confidence level. Fluxes are $\ergpcmsqps$ Flux ratio is the \texttt{bb} flux divided by the \texttt{diskbb} flux. Luminosities are in units of $10^{38}\ergps$. }
\label{tab:T1}
\begin{tabular}{ccccccc}
\hline\hline
Observation &  & 23 & 25 & 27 & 29 & 31 \\
\hline
Total Model & Flux ($0.1 -78\keV$) & 1.62e-7 & 1.30e-7 & 1.03e-7 & 7.81e-8 & 6.98e-8 \\
  & Luminosity  & 2.26  & 1.8 & 1.43 & 1.08 & 0.967 \\
\hline
\texttt{diskbb} & $kT_{\rm in}\keV$ & $0.603\pm{0.01}$ & $0.650\pm{0.12}$ & $0.659\pm0.01$ & $0.651\pm0.013$ & $0.644\pm0.007$ \\
	& Norm & 4.45e4$^{+5{\rm e}3}_{-4.5{\rm e}3}$ & 3.02e4$\pm{3{\rm e}3}$ & 2.35e4$^{+1.8{\rm e}3}_{-1.6{\rm e}3}$ & 1.94e4$^{+2.1{\rm e} 3}_{-1.7{\rm e}3}$ & 1.82 e4$^{+1.1{\rm e}3}_{-9.7{\rm e}2}$ \\ 
	& Flux ($0.1-78 \keV$) & 1.24e-7 & 1.14e-7 & 9.39e-8 & 7.35e-8 & 6.63e-8\\
\hline
\texttt{bbody} & $kT\keV$ & $1.12\pm{0.012}$ & $0.93\pm{0.02}$ & $0.87\pm0.014$ & $0.89\pm0.03$ & $0.92\pm0.02$\\
	& Norm & $0.177\pm{0.004}$ & $0.116\pm{0.012}$ & $0.087\pm0.01$ & $0.045\pm0.009$ & $0.03\pm0.005$\\
	& Flux ($0.1-78 \keV$) & 1.48e-8 & 9.76e-9 & 7.28e-9 & 3.74e-9 & 2.54e-9 \\
	& Area (1e12\cmsq) & 12.7 & 17.6 & 17.4 & 8.1 & 4.8 \\
	& $\Delta R \km $ & 3.36 & 4.66 & 4.61 & 2.14 & 1.27 \\
	& Flux Ratio & 0.12 & 0.086 & 0.078 & 0.05 & 0.038 \\
	& Temp Ratio & 1.86 & 1.43 & 1.31 & 1.25 & 1.42\\
\hline
\texttt{cutoffpl} & $\Gamma$ & $1.6\pm{0.06}$ & $1.47\pm0.07$ & $1.45\pm0.063$ & $1.92\pm0.33$ & $2.14\pm0.24$  \\
                 & $E_{\rm cut}\keV$ & $49.8\pm{5.5}$ & $73.8^{+18}_{-12}$ & $74^{+17}_{-12}$ & $36.9^{+43}_{-14}$ & $64.6^{+114}_{-26}$ \\
                 & Norm & $1.67\pm{0.2}$ & $0.284\pm0.04$ & $0.093\pm0.012$ & $0.083^{+0.09}_{-0.05}$ &  $0.116^{+0.07}_{+0.04}$  \\
   	& Flux ($0.1-78 \keV$) & 2.39e-8 & 6.0e-9 & 2.06e-9 & 7.97e-10 & 1.0e-9 \\  
	  & Flux ($1-78 \keV$) & 1.99e-8 & 5.4e-9 & 1.86e-9 & 5.0e-10 & 5.0e-10 \\      
\hline\hline
\texttt{ezdiskbb} & $kT_{\rm max}\keV$ & $0.574\pm{0.01}$ & $0.615\pm{0.11}$ & $0.621\pm0.01$ & $0.615\pm0.013$ & $0.615\pm0.007$ \\
	& Norm & 9.39e3$^{+1.0{\rm e}3}_{-9.3{\rm e}2}$ &6.6e3$^{+7{\rm e}2}_{-6{\rm e}2}$ & 5.2e3$^{+3.9{\rm e}2}_{-3.4{\rm e}2}$ & 4.24e3$^{+4.6{\rm e} 2}_{-3.7{\rm e}2}$ & 3.95e3$^{+2.4{\rm e}2}_{-2.1{\rm e}2}$ \\ 
\hline
\texttt{bbody} & $kT\keV$ & $1.12\pm{0.012}$ & $0.92\pm{0.02}$ & $0.86\pm0.014$ & $0.88\pm0.03$ & $0.91\pm0.02$\\
	& Norm & $0.177\pm{0.004}$ & $0.122\pm{0.11}$ & $0.095\pm0.01$ & $0.049\pm0.01$ & $0.033\pm0.004$\\
\hline
\end{tabular} 
\end{table*}

 \begin{table*}\centering
  \caption{The best-fit model parameters for the model
    \texttt{kerrbb+bb+cutoffpl}. }
    \label{tab:T2}    
\begin{tabular}{ccccccc}
\hline\hline
Observation & & 23 & 25 & 27 & 29 & 31 \\
\hline
\texttt{kerrbb} & $Mdd$ &  $1.72\pm{0.13}$ & $1.92\pm0.009 $ & $1.70\pm0.006$ & $1.37\pm0.009$ & $1.25\pm0.005$ \\
  	& $F (0.1-78 \keV$) & 9.43e-8 & 1.045e-7 & 9.23e-8 & 7.45e-8 & 6.81e-8 \\   	
	\hline
\texttt{bbody} & $kT\keV$ & $1.24\pm{0.01}$ & $1.08\pm{0.015}$ & $0.953\pm0.009$ & $0.937\pm0.016$ & $0.96\pm0.01$ \\
	& Norm & $0.138\pm{0.002}$ & 5e-2$\pm$0.17e-3 & 4e-2$\pm$1.5e-3 & 2.9e-2$\pm$1.7e-3 & 2.15e-2$\pm$8e-4 \\
  	& $F(0.1-78 \keV$) & 1.19e-8 & 5.06e-9 & 4.36 e-9 & 3.07e-9 & 2.29e-9 \\
	& Flux Ratio & 0.126 & 0.048 & 0.047 & 0.041 & 0.033 \\
\hline
\end{tabular} 
\end{table*}

\subsubsection{Results from {\em  NuSTAR} spectral fits}
Nu21 to 31 have been jointly fitted  with the spectral models  \texttt{diskbb+bb+cutoffpl},   \texttt{ezdiskbb+bb+cutoffpl} and  \texttt{kerrbb+bb+cutoffpl}. \texttt{diskbb} and \texttt{ezdiskbb}  are blackbody accretion disk models, the first (Makishima et al. 1986) assumes continuing torque to the inner radius and the second assumes zero torque there (Zimmerman et al. 2005).  For \texttt{kerrbb} the spin $a$ and inclination $i$ have been fixed at 0.2 and 34 deg respectively, based on the hard state results (Buisson et al. (2019). $Mdd$ is the mass accretion rate in units of $10^{18}\gps$. $\chi^2$ for the 3 models (\texttt{diskbb, ezdiskbb, kerrbb}) were 8096, 8079 and  8401 for 6944, 6984 and 6945 degrees of freedom, respectively. 

The properties of the excess blackbody components are given in \ref{tab:T1} and \ref{tab:T2}.     
Where the disc blackbody is used, the surface area required at the source, assumed to lie at $3.5\kpc$, has been estimated\footnote{Relativistic blurring of a ring at $5\rg$ can lead to systematic changes in the estimated temperature of 5--10 per cent, depending on inclination.}. Assuming that the area is that of a thin ring at the Innermost Stable Circular Orbit (ISCO) lying at $5r_{\rm g}$  (Buisson et al. 2019), 60\,km from an $8\Msun$ black hole, then we find a width $\Delta R$ ranging  4.66 ($0.4 r_{\rm g}$, Nu25) to 1.27\km ($0.1 r_{\rm g}$, Nu31) which  will systematically vary as $\cos i$. The flux in the excess blackbody compared with that of the disc varies monotonically from 12 to 3.8 per cent from Nu23 to Nu31 when \texttt{diskbb} is used and 12.6 to 3.3 per cent for \texttt{kerrbb}, mostly due to changes in  normalization rather than temperature. The flux in the powerlaw component depends on the low energy limit assumed: for $0.1\keV$ it drops by about a factor of 40 in flux from 16 per cent of \texttt{diskbb} to 0.8 per cent. This disparity between the flux changes  of the excess blackbody and powerlaw components argues against any reflection interpretation (see later). The extent of the variations in the blackbody do not obviously associate it directly with either the disc or the power-law component.

\begin{figure}
    \includegraphics[width=1.\columnwidth]{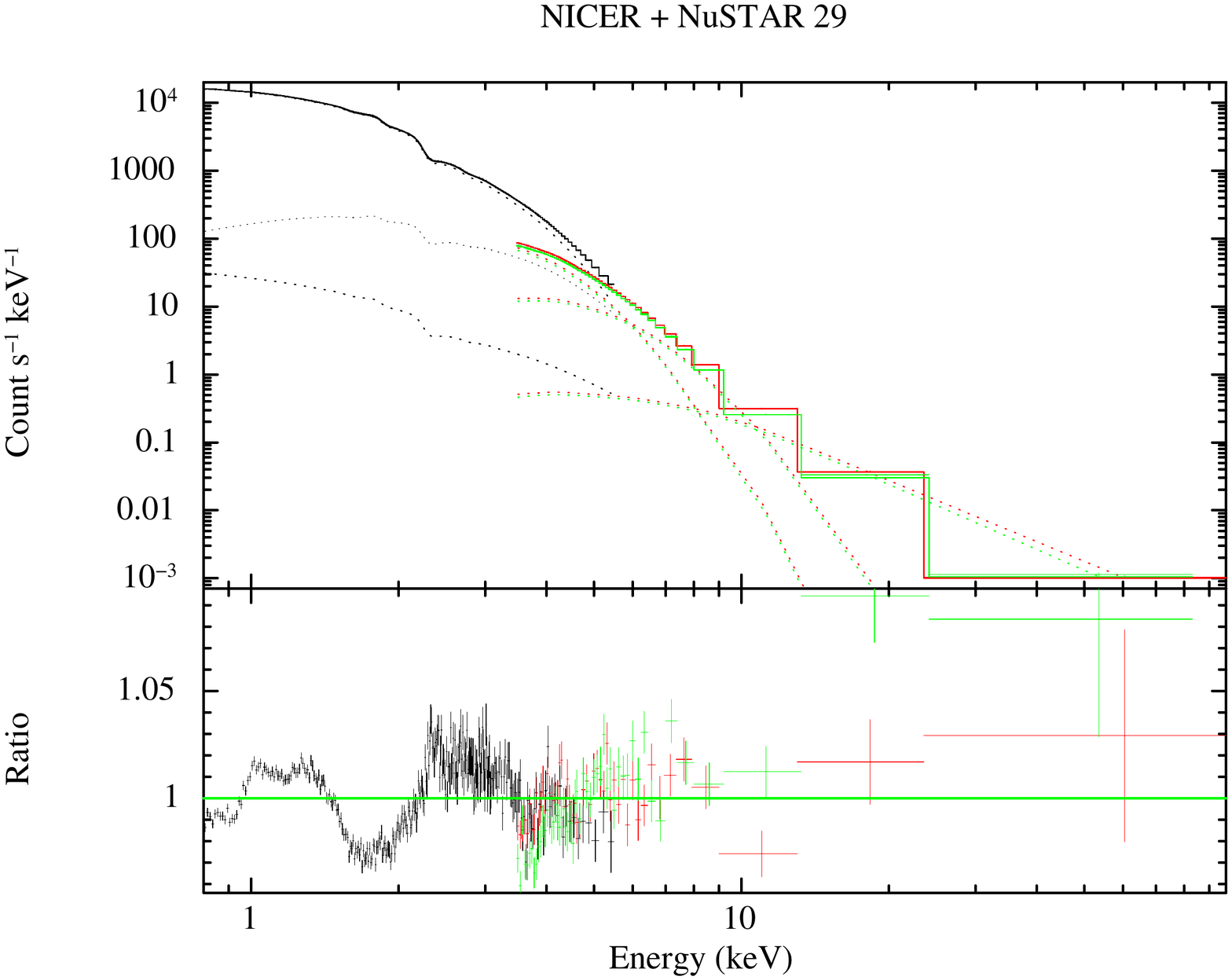}  
     \includegraphics[width=1.\columnwidth]{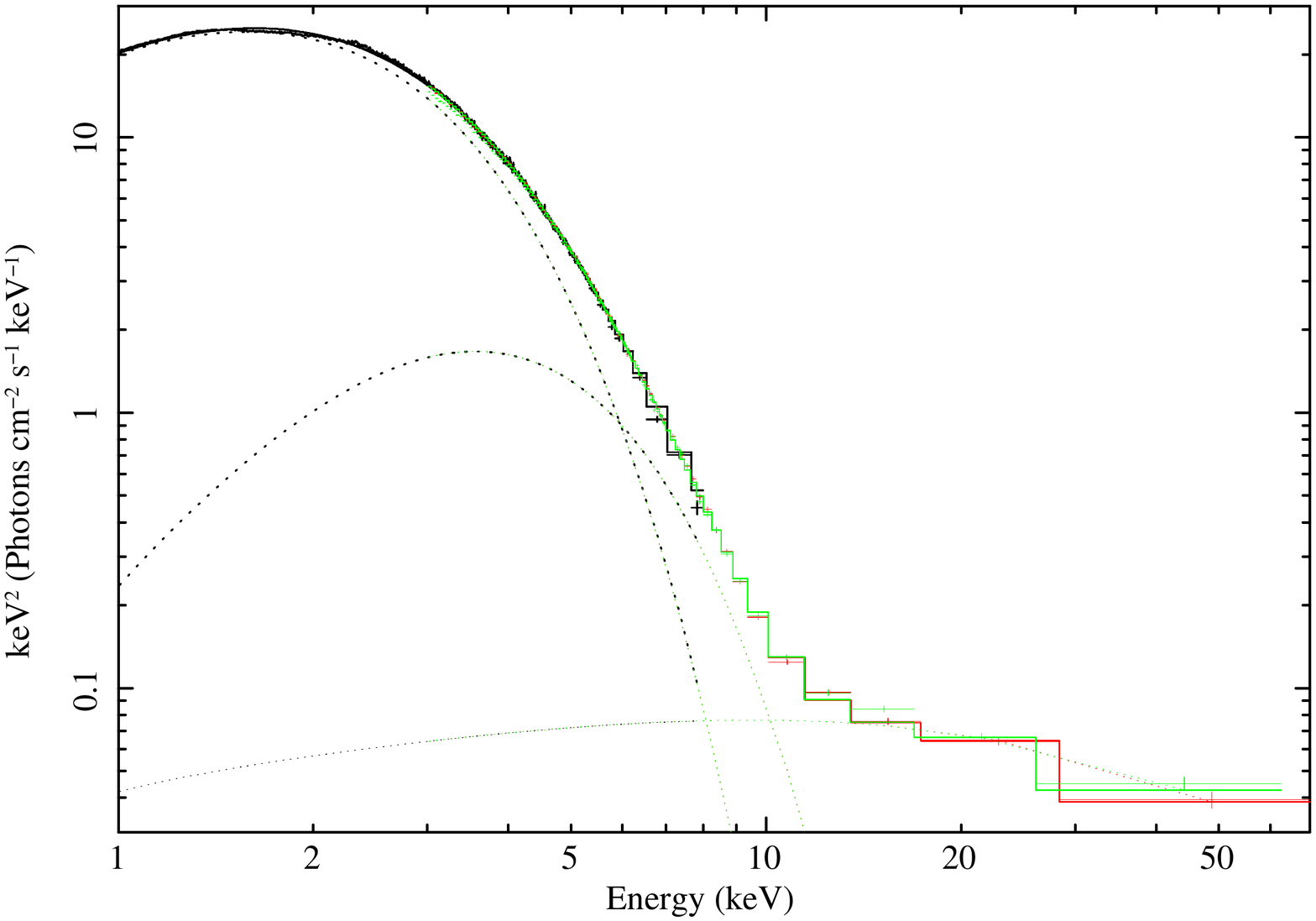}    
   \caption{Joint {\em NICER} and {\em  NuSTAR} fit to epoch 29 with \texttt{constant*(ezdiskbb + bb +cutoffpl)}. The top plot shows the fit in counts and the data-model ratio, the lower plot shows the unfolded spectrum with model components dashed. } 
\label{fig:ninu}
\end{figure}

In \ref{fig:ninu} we show a joint fit of \texttt{constant * (ezdiskbb + bb + cutoffpl)} to the {\em NICER} and {\em  NuSTAR} data from epoch 29.  Again, the fit is acceptable in a $\chi^2$ sense if 1.5 per cent systematics are included. The fit confirms that the \texttt{bb} component has a higher temperature than that of \texttt{ezdiskbb}.  (This issue was ambiguous when just {\em  NuSTAR} is used.)  While \texttt{constant} is set to unity for {\em  NuSTAR} module A, it fits to 0.43 for {\em NICER} due to over half the mirror modules being switched off to deal with  the high count rate. The parameters for \texttt{ezdiskbb} and \texttt{bb} are similar to those listed in Table 1, the powerlaw fit is changed to $\Gamma=1.6\pm0.33,\, E_{\rm cut}=25^{+17}_{-8}\keV$ and $Norm=5e-2$. 

Repeating the above fit with \texttt{kerrbb} replacing \texttt{ezdiskbb}, we have explored parameter space for spin and inclination. The results depend on the detailed treatment of the power-law component but using the above $\Gamma$ and $E_{\rm cut}$ values we find $a=0.64^{+0.2}_{-0.4}$ and $i=21.6^{+2.0}_{-1.5}$. The dependence of $a$ on  $i$ is  similar to that shown in the lower panel of Fig. 2, but shifted to lower inclinations by about 7 deg. The black hole mass is determined as about $7.7\Msun$.

We have also considered whether the 7--9 keV excess can be due to iron emission (and absorption) as in the models by Ross \& Fabian (2007).  These model  a slab illuminated from above by a powerlaw and from below by a blackbody, thereby mimicking an irradiated  blackbody disc. Strong iron K features are seen when the blackbody has a temperature of 0.3 keV, but the features are weak by a temperature of 0.5 keV (Fig. 6 of Ross \& Fabian 2007, which shows just the radiation emerging from the slab).   First, in order to measure the amplitude of any Fe-K emission, we replaced the extra blackbody with a gaussian component, centred at 6.5\,keV. This gives a reasonable fit to the joint data of Nu29 but for a standard deviation of about 1.5 keV. Under this model the excess shown in Fig.~\ref{fig:nu29rat} represents  the high energy shoulder of the gaussian component. The equivalent width (EW) of the gaussian emission is almost 4\,keV when measured against the powerlaw component alone, which we consider too strong to be plausible (see Matt, Fabian \& Ross 1993 and Zycki \& Czerny 1994 for iron K EW predictions).

Second we tried fitting the \texttt{refbhbhi} model from Ross \& Fabian (2007), which involves the whole reflection spectrum,  to the joint Nu29 data. We find a reasonable fit for  an inclination of 85 deg with an uncertainty of less than 1 deg. Cstat increases by about 40 if reduced to 77 deg in order to become consistent with the optical data. This result could imply that reflection may be involved but is not a clear answer. A major concern is that Fe-K reflection features are produced by  irradiation of the disk by photons near 7\,keV. The flux from the power law component at that energy varies by a factor of 41 from Nu23 to Nu31, whereas the flux of the excess component, measured as the extra blackbody, varies only by a factor of only 5.9. Unless the geometry of the irradiation changes by significant factors then we do not see how they are causally connected. 

\begin{figure} 
  \includegraphics[width=\columnwidth]{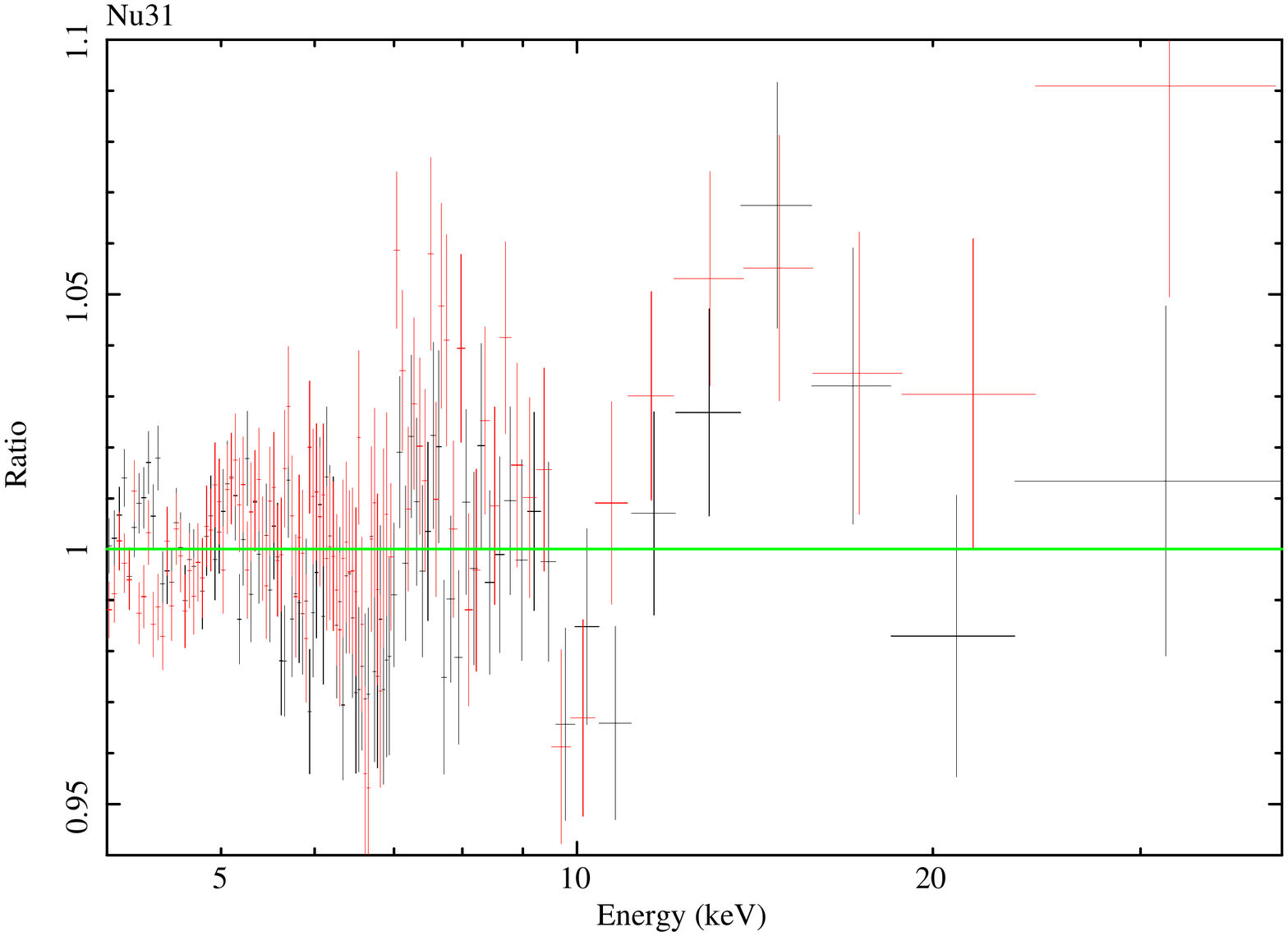}
  \includegraphics[width=\columnwidth]{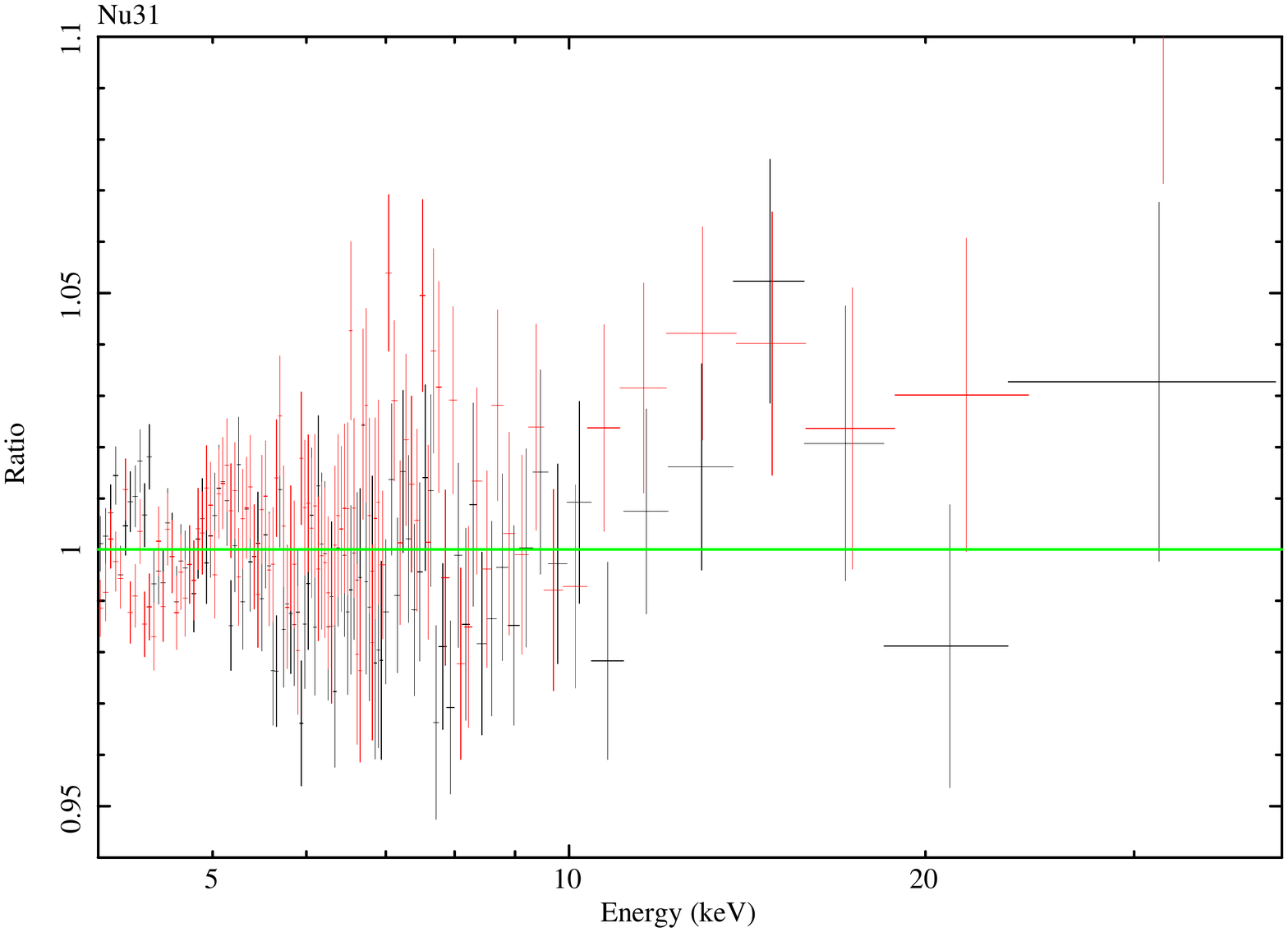}
\caption{Top: Ratio of Nu31 to best-fit model with Kerrbb and 3bb. Bottom: as above but after gaussian absorption at 6.6 keV and edge at 9.2 keV have been included in the fitted spectrum.}
\label{fig:nu31bb_nu31edge}
\end{figure}

One characteristic of the spectral residuals seen in \ref{fig:nu2331dbb_kbb} is a broad dip at 10 keV and some spectra (Nu27 and Nu31) also show a dip just below 7 keV. For the latter, ionized absorption is a likely explanation. Therefore we have added a negative gaussian at around 6.6 keV and an edge at about 9.2 keV. This improves the spectral fit by $\Delta\chi^2=-35$ (Fig.~\ref{fig:nu31bb_nu31edge}). A separate fit with SPEX using the model \texttt{pion} reveals an ouflowing wind with column density of $2-4\times10^{21}\psqcm$ and outflow velocity of around $4000\kmps$ and ionization parameter $\xi\sim 3000\erg\,\cm^{-2}\,\s^{-1}$. The fit is not improved significantly by using that model for the earlier observations Nu23. 25 or 29. 
 We note that such outflows are common in the soft state of black hole X-ray binaries (Ponti et al. 2012), usually if they have high inclination.  
 
\section{The Plunge Region}

The additional component seen in the spectra is well explained by the addition of another blackbody, with a temperature slightly higher than that of the disc itself. The ratio of temperature of the blackbody to the inner  \texttt{diskbb} temperature is between 1.25 (Epoch 27) and 1.86 (Epoch 23, see Table 1). It appears to be a separate component. Since temperatures increase inward in the disc we associate it with the innermost part of the disc and with the start of the plunge region.

\begin{figure}
\includegraphics[width=1.\columnwidth]{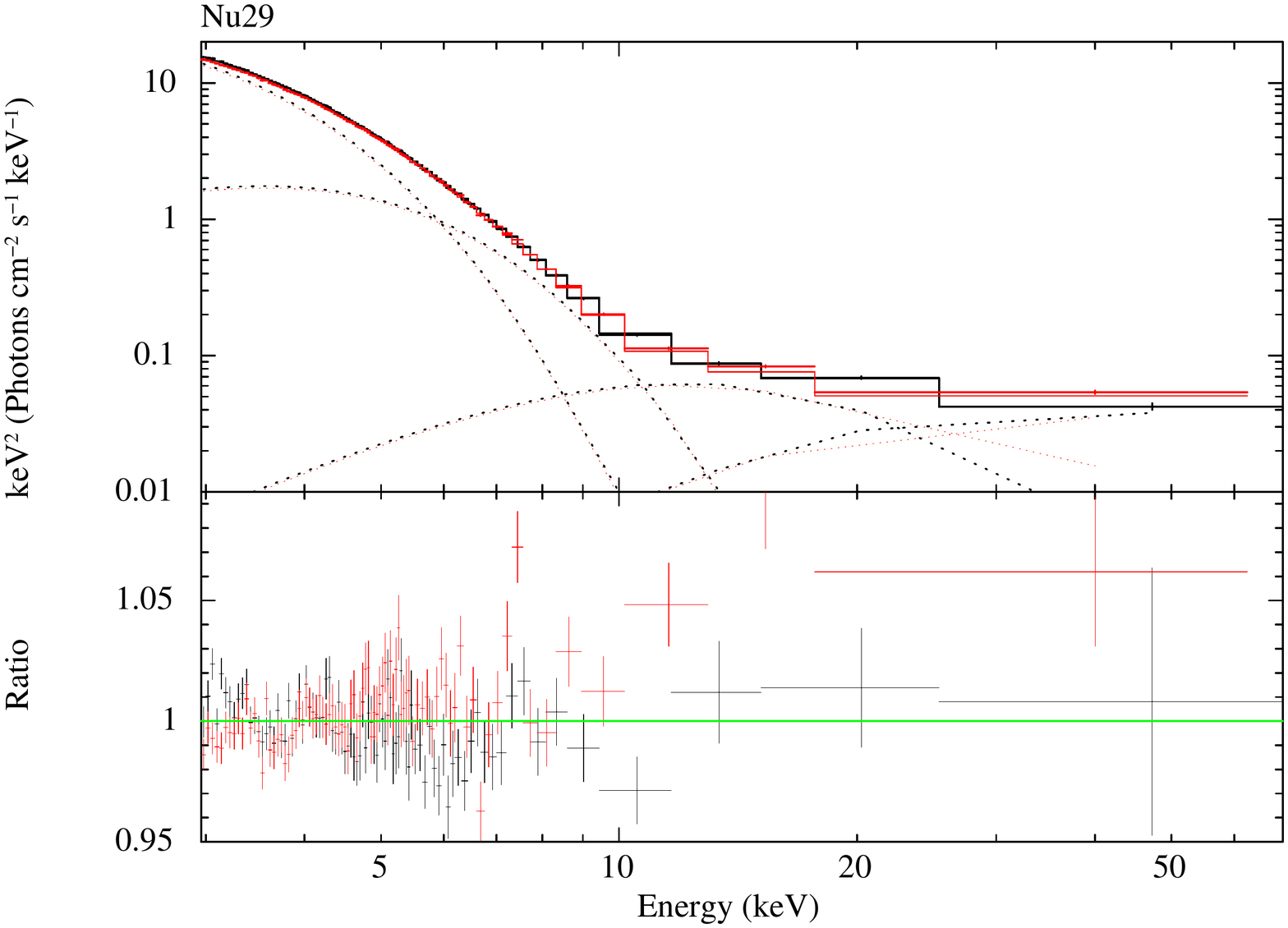}
\caption{{\em NuSTAR} spectra of observation 29  with the hard tail fitted by  2 extra blackbody components. }
\label{fig:nu293bb}
\end{figure}

The standard Shakura \& Sunyaev (1973) accretion disc assumes a zero-stress boundary condition at the Innermost Stable Circular Orbit (ISCO). Matter enters ballistic plunge orbits at that radius with very little dissipation or further emission as it falls into the black hole. Later work in which the effect of magnetic fields are considered has suggested that stresses can and do occur at the ISCO leading to dissipation and  excess emission as matter enters the plunge region (Reynolds \& Armitage 2001, Hawley \& Krolik 2002, Machida \& Matsumoto 2003, Shafee et al. 2008, Zhu et al. 2012, and Abolmasov 2014).   One of the most detailed studies, by Zhu et al. (2012) shows that further inward flow can cause a weak power-law-like continuum composed of  ever hotter black body (bb) emission as matter falls further. Modelling the {\em  NuSTAR} spectra as a disc blackbody together with 3 to 4 blackbodies \texttt{diskbb + bb + bb + bb}, does give reasonable fits (e.g. Fig.~\ref{fig:nu293bb}).

\begin{figure}
  \includegraphics[width=\columnwidth]{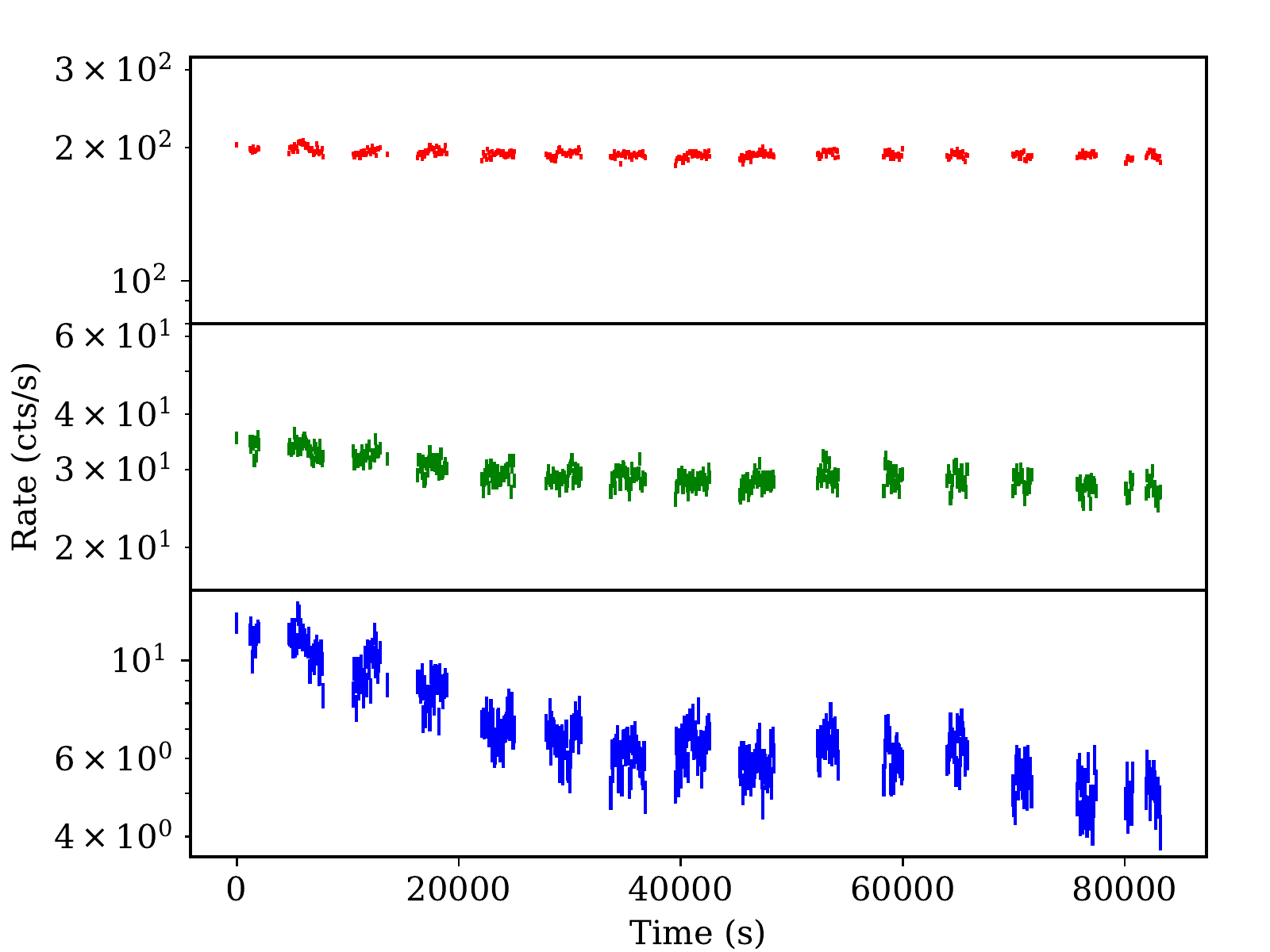}
  \includegraphics[width=\columnwidth]{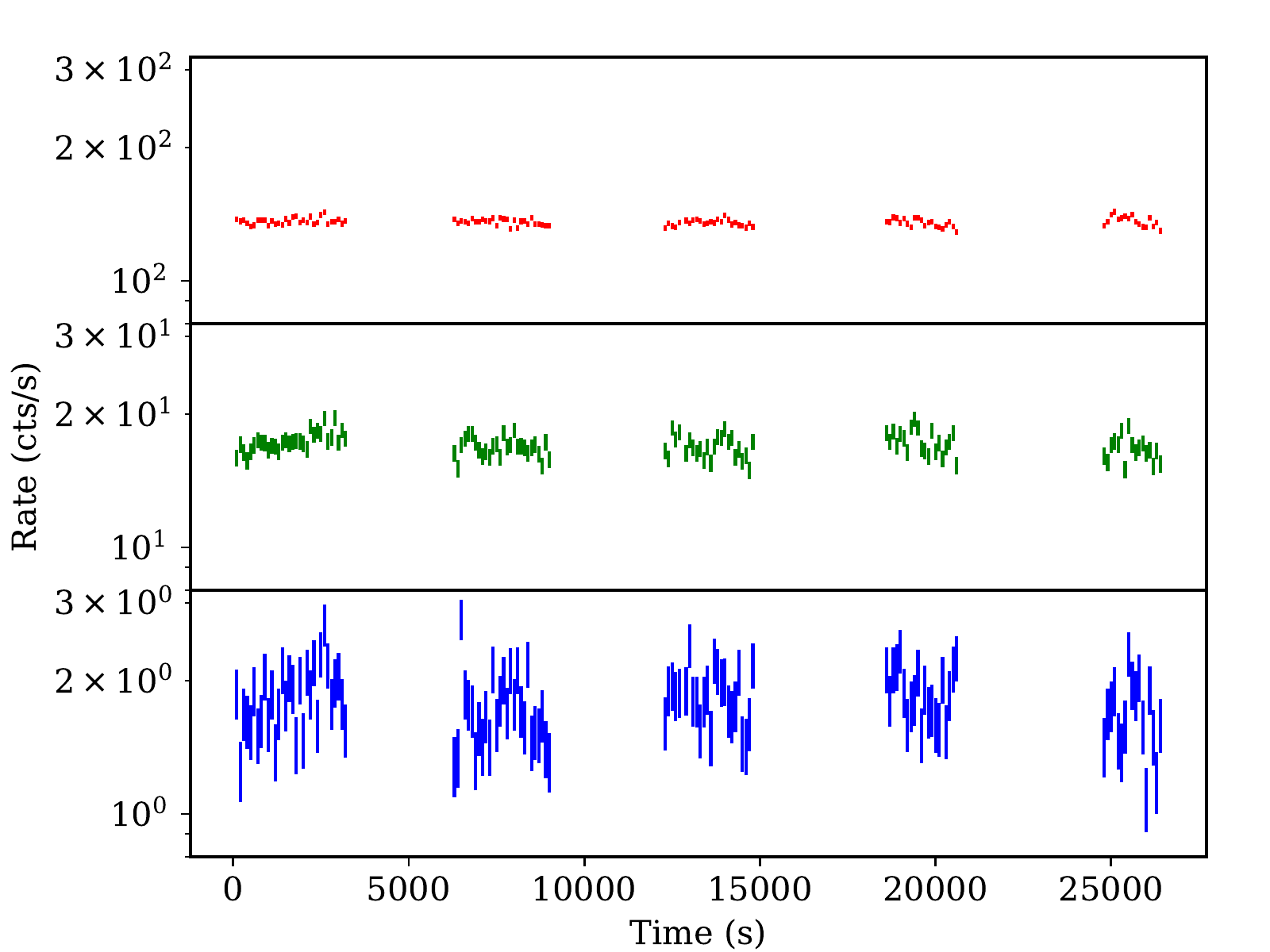}
 \caption{Top:  {\em NuSTAR} lightcurves of observation 27 (above) and 29 (below) in the energy bands 3--4, 6--8 and 10--78 keV (top to bottom). The bin size is 100 s.}
\label{fig:lc27_29}
 \end{figure}

 \begin{figure} 
  \includegraphics[width=\columnwidth]{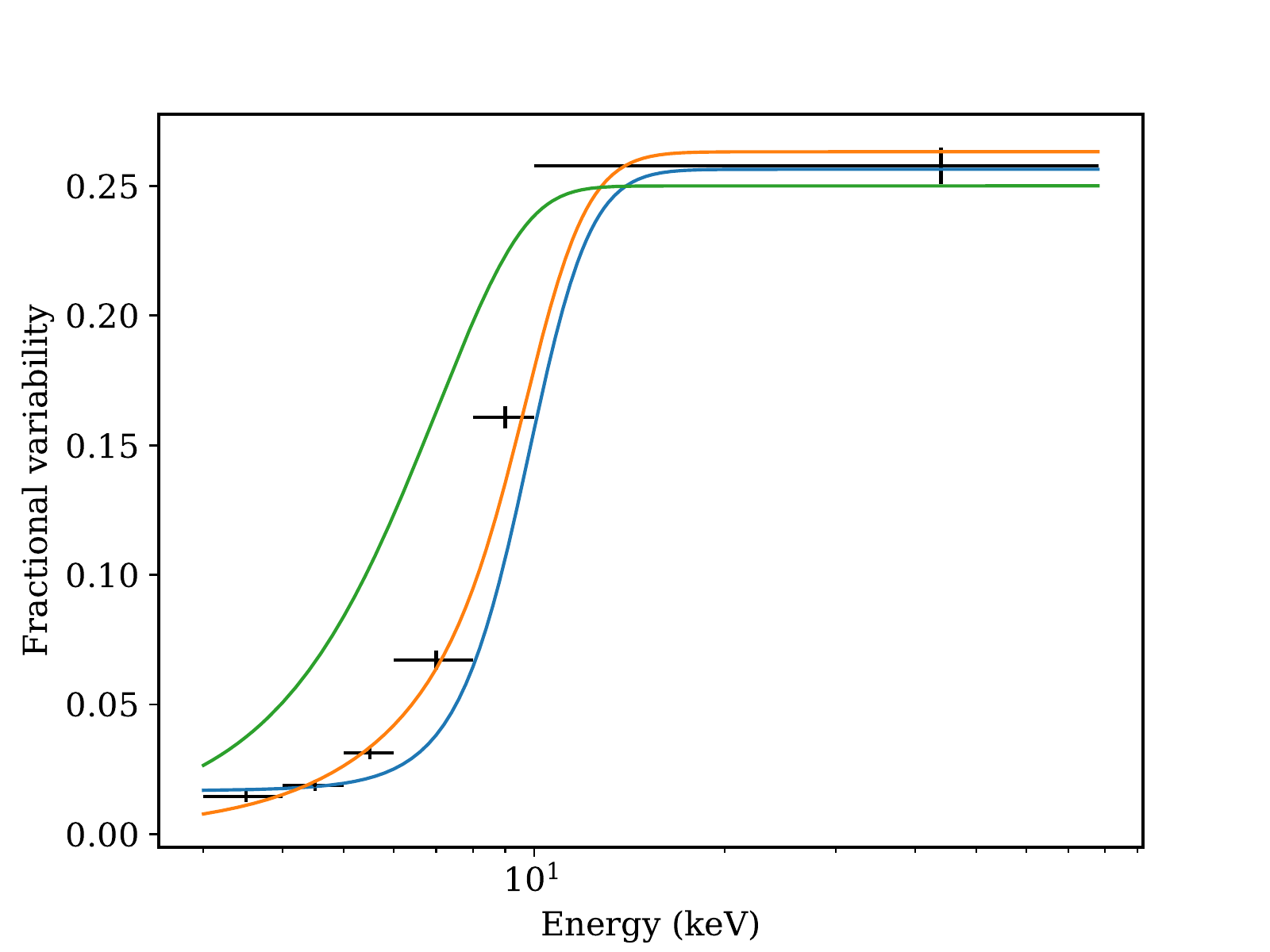}
  \includegraphics[width=\columnwidth]{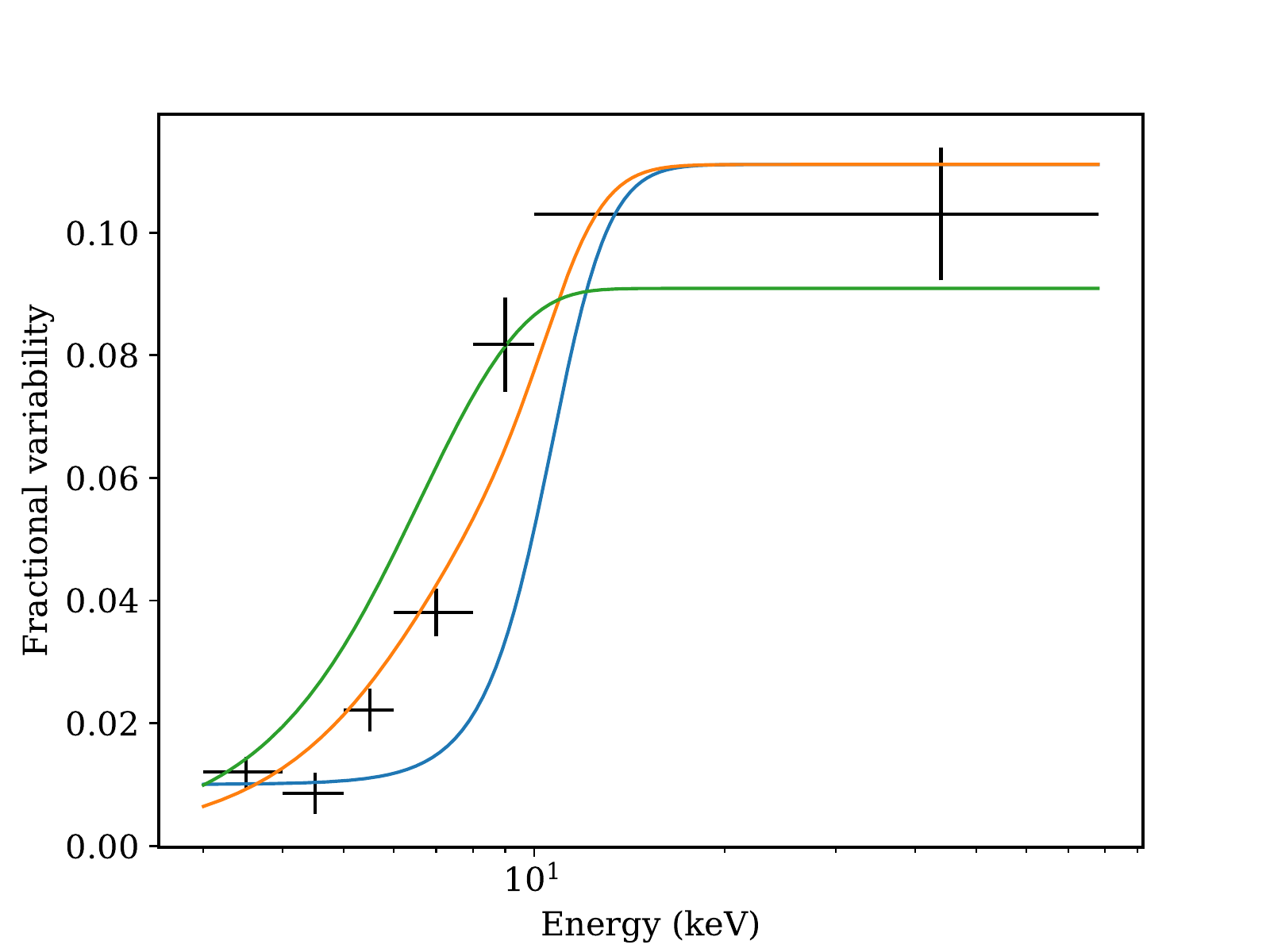}
\caption{ Fractional variability in observations 27 and 29.  The solid lines show the expectation when the additional blackbody has the same amplitude as the disc (lower line), when the amplitude is freely fitted (middle line) and when  it has the same amplitude as the power-law component (upper line).  }
\label{fig:vb27_29}
\end{figure}

Much as a waterfall produces chaos from a smooth stream, the flow from
an accretion disc into the plunge region is expected to be highly
variable (Hawley \& Krolik 2002, Machida \& Matsumoto 2003). We indeed find that the general disc emission varies little
but the emission above 6 keV is increasingly variable (Fig.~\ref{fig:lc27_29}, Fig.~\ref{fig:vb27_29}). For Nu27 the power-law component shows 25 per cent variability (including a significant drop throughout the observation) and for Nu29 it is 10 per cent.  The rapid variability shown in the latter drops across the 10 to 6 keV band in a manner midway between that of the power-law component and the disc. 
Further work on the variability is beyond the scope of this paper and will be reported elsewhere.

\section{Discussion}

MAXI J1820+070 has been well observed by {\em NuSTAR} and {\em NICER} during the soft state of its bright outburst in 2018. The source peaked at an Eddington fraction of about 20 per cent before declining to about 10 per cent (assuming a blackhole mass of $8\Msun$). Here we have examined the soft-state spectra and shown that in addition to the usual disk-blackbody and power-law components, a further blackbody component is required. The flux of this extra component is about 12 per cent in the soft intermediate state (Nu23) and then declines during the full soft state (Nu25-31) from about 4.8 to 3.3 percent of the flux of the disk component.
A similar excess blackbody is required if a \texttt{kerrbb} model is used instead of \texttt{diskbb}. 

A straightforward explanation for the excess is emission from the innermost part of the accretion flow across the ISCO, as the accreting matter begins to fall in the plunge region and thence to the black hole.  Magnetic stresses at the ISCO can delay the infall, increasing the gravitational energy release which powers the emission. The situation is complex and poorly understood and no clear predictions are available. Modelling the luminosity and temperature of the excess blackbody through simulations should lead to further understanding of the strength and nature of the magnetic fields  around the ISCO and their evolution with time. 

Reflection could also be involved, but is disfavoured as a major contributor due to the large difference in the rate of change of the powerlaw component and the 6--10 keV  excess. Any such reflection would also originate from the innermost part of the disc.

Zhu et al. (2012) show that the Thomson optical depth through the disc remains high within the beginning of the plunge region, so radiation has to diffuse out. Moreover, despite dropping inward, the density remains high enough for the electrons and protons in the gas to be thermalized,  so blackbody radiation is expected.

The effect of non-zero torque can be mimicked in \texttt{kerrbb} with the torque parameter $\eta$. Normally this is set to zero, as has been the case so far here. If we allow $\eta$ to be a free parameter and just fit $\texttt{constant}*(\texttt{kerrbb}+\texttt{cutoffpl})$ to the {\em  NuSTAR} and {\em NICER} data of epoch 29, then the fit improves. It reaches a $\chi^2$ minimum at about 400 more than when \texttt{bb} is used, but does halve the residuals in the 6--10\,keV band. On discussion of \texttt{kerrbb}, Li et al. (2005) point out that the use of $\eta$ will only be partially successful since the model assumes no emission within the ISCO, which is physically unlikely.  The net result is to support our claim that the blackbody excess is associated with a non-zero torque  leading to excess emission both immediately outside and {\it inside} the ISCO.

The power-law component is not simple but has curvature which we model with a cutoff. This improves the fit but leaves significant residuals.  The component shows large variability of 10 to 25 per cent. The excess blackbody component shows variability intermediate between that of the power-law and the disk components (as expected, the disc itself shows little variability). The powerlaw component can also be modelled with a sequence of blackbody spectra (Fig.~\ref{fig:nu293bb})) which could originate in the plunging flow (see Zhu et al. 2012).  

One question remains:  why is the $6-10\keV$  spectral component not commonly seen? First we note that the effect may be small when the black hole spin is high and the plunge region relatively small due to the proximity of the ISCO to the event horizon of the black hole (see discussion by Zhu et al. 2012).   Second,   
the effect is quite subtle, requiring accurate measurement of an excess flux of only 10 per cent in the 7-12 keV band. (Figs. 4 and 5).  Thirdly, a few sources do sometimes show high temperature components requiring a small emitting region (e.g. Muno et al. 2001;  Tomsick et al. 2005) although the overall situation of those sources is quite different from that studied here. Finally, it may also become confused if there is a significant reflection component.

We note that Oda et al. (2019) have reported a similar excess in single epoch {\it Suzaku} spectra of  the black hole candidate MAXI\,J1828-249 taken during the intermediate state as the soft state was beginning. They explain it using separate Comptonization components, one of which  accounts for  the power-law tail and the other the 5--10\, keV excess. The latter is due to photons from either a)  the inner edge of the \texttt{diskbb}, which has a temperature similar to that seen in MAXI\,J1820+070,  upscattered using \texttt{nthcomp} in an electron cloud of temperature $1.2\keV$ and optical depth $\tau>3.5$, or b) the whole disc with a cloud of temperature 14\,keV and optical depth about unity.   We find that model a) gives a good fit to the epoch 29 {\em NICER} and {\em  NuSTAR} data with $\tau\sim25$ but is {\em very} sensitive to the electron temperature ($kT_{\rm e}= 0.905^{+0.016}_{-0.021}\keV$). A visual comparison of the best-fitting \texttt{nthcomp} and \texttt{bb} spectra shows them to be almost indistinguishable above 3 keV, so from the NuSTAR point of view of fitting MAXI\,J1820+070 they can be considered equivalent. We therefore suspect that blackbody emission from the ISCO is also relevant to MAXI\,J1828-249.

Earlier, work on the high state of GX\,339-4 by Kotehlainen, Done \& Diaz Trigo (2011) shows a continuum component that is broader than disc blackbody models. The breadth was ascribed to an additional Comptonization component, which in this case has a luminosity nearly comparable to that of the \texttt{diskbb} component. No physical explanation of the component is given and $\tau$ is so large that its shape is again close to that of a blackbody. A significant reflection component is also present, unlike the case presented here for MAXI J1820+070. 

We conclude that an excess blackbody component is the simplest explanation for the soft state spectral behaviour of MAXI J1920+070 and identify that component with the start of the plunge region.

\section*{Acknowledgements}
ACF and SD acknowledge support from ERC Advanced Grant FEEDBACK, DRW is supported by NASA through Einstein Postdoctoral Fellowship grant number PF6-170160, awarded by the \textit{Chandra} X-ray Center, operated by the Smithsonian Astrophysical Observatory for NASA under contract NAS8-03060.340442. C.S.R. thanks the UK Science and Technology Facilities Council (STFC) for support under the New Applicant grant ST/R000867/1, and the European Research Council (ERC) for support under the European Union's Horizon 2020 research and innovation programme (grant 834203).



\begin{thebibliography}{99}

\bibitem[\protect\citeauthoryear{Abolmasov}{2014}]{Abolmasov2014}
 Abolmasov P., 2014, MNRAS, 445, 1269
 \bibitem[\protect\citeauthoryear{Bright}{2019}]{Bright2019}
 Bright J., Motta S., Fender R., Perrott Y., Titterington D., 2018, The Astronomer's Telegram 11827
\bibitem[\protect\citeauthoryear{Buisson}{2019}]{Buisson2019}
Buisson D.J., et al. 2019,  MNRAS, 490, 1350
\bibitem[\protect\citeauthoryear{Gandhi}{2019}]{Gandhi2019}
Gandhi P., Rao A., Johnson M.A.C., Paice J.A., Maccarone T.,  2019. MNRAS 485, 2642
\bibitem[\protect\citeauthoryear{Gendreau}{2016}]{Gendreau2016}
Gendreau K.C., et al. 2016, SPIE, 9905, 16
\bibitem[\protect\citeauthoryear{Hawley}{2002}]{Hawley2002}
Hawley J.F., Krolik, J.H.,  2002, ApJ, 566, 164
\bibitem[\protect\citeauthoryear{Kara}{2019}]{Kara2019}
Kara E. et al.,  2019, Nature, 565, 198
\bibitem[\protect\citeauthoryear{Kawamuro}{2018}]{Kawamuro2018}
Kawamuro T, et al.,  2018, The Astronomers Telegram, 11399,1 
\bibitem[\protect\citeauthoryear{Li}{2005}]{Li2005}
Li L., Zimmerman E.R., Narayan R., McClintock J.E. 2005, ApJS, 157, 335
\bibitem[\protect\citeauthoryear{Kotehlainen}{2011}]{Kotehlainen2011}
Kotehlainen M., Done C., Diaz Trigo M., 2011, MNRAS, 416, 311
\bibitem[\protect\citeauthoryear{Ludlam}{2018}]{Ludlam2018}
Ludlam R., et al., 2018, ApJ, 858, L5
\bibitem[\protect\citeauthoryear{Machida}{2003}]{Machida2003}
Machida M, Matsumoto R., 2003, ApJ, 585, 429
\bibitem[\protect\citeauthoryear{Makishima}{1986}]{Makishima1986}
Makishima K., et al., 1986, ApJ, 308, 635
\bibitem[\protect\citeauthoryear{Matsuoka}{2009}]{Matsuoka2009}
Matsuoka M. et al., 2009. PASJ, 69, 999
\bibitem[\protect\citeauthoryear{Matt}{1993}]{Matt1993}
Matt G., Fabian A.C., Ross R.R. 1993, MNRAS, 262 179
\bibitem[\protect\citeauthoryear{Muno}{2001}]{Muno2001}
Muno M. et al., 2001, ApJ, 556. 515
\bibitem[\protect\citeauthoryear{Oda}{2019}]{Oda2019}
Oda, S. et al., 2019, PASJ, 71, 108
\bibitem[\protect\citeauthoryear{Ponti}{2012}]{Ponti2012}
Ponti G., Fender R. P., Begelman M. C., Dunn R. J. H., Neilsen J., Coriat M. 2012, MNRAS, 422, L11
\bibitem[\protect\citeauthoryear{Reynolds}{2001}]{Reynolds2001}
Reynolds C.S., Armitage P.J., 2001, ApJ, 561, L81
\bibitem[\protect\citeauthoryear{Ross}{2007}]{Ross2007}
Ross, R.R., Fabian, A.C., 2007, MNRAS, 381, 1697
\bibitem[\protect\citeauthoryear{Shakura}{1973}]{Shakura1973}
Shakura N.I., Sunyaev R.A., 1973, A\&A, 24, 337
\bibitem[\protect\citeauthoryear{Shidatsu}{2019}]{Shidatsu2019}
Shidatsu M., et al., 2019, ApJ, 874, 173
\bibitem[\protect\citeauthoryear{Steiner}{2009}]{Steiner2009}
Steiner J.F., Narayan R., McClintock J.E., Ebisawa K., 2009, PASJ, 121, 1279
\bibitem[\protect\citeauthoryear{Tomsick}{2005}]{Tomsick2005}
Tomsick J.A., Corbel S., Goldwurm A, Kaaret P., 2005, ApJ, 630, 413
\bibitem[\protect\citeauthoryear{Torres}{2019}]{Torres2019}
Torres M.A.P., Casares J., Jimenez-Ibarra, F., Munoz-Darias, T., Armas Padilla M., Jonker, P.G., Heida M., 2019,  ApJL, 882, L21
\bibitem[\protect\citeauthoryear{Tucker}{2018}]{Tucker2018}
Tucker M. et al., 2018, ApJ, 867, L9
\bibitem[\protect\citeauthoryear{Zhu}{2012}]{Zhu2012}
Zhu Y., Davis, S.W., Narayan R., Kulkarni, A.K., Penna, R.F., McClintock J.E., 2012, MNRAS, 424, 2504
\bibitem[\protect\citeauthoryear{Zimmerman}{2005}]{Zimmerman2005}
Zimmerman E., Narayan R., McClintock J.E., Miller J.M.,  2005, ApJ, 618, 832
\bibitem[\protect\citeauthoryear{Zycki}{1994}]{Zycki1994}
Zycki P. Czerny  B., 1994, MNRAS, 266 653

\end{thebibliography}

\bsp	
\label{lastpage}
\end{document}